\documentclass[amssymb,amsfonts,preprint,floatfix,showpacs,superscriptaddress,10pt]{revtex4-1}

\usepackage[utf8]{inputenc}
\usepackage{orcidlink}
\usepackage[french,english]{babel}
\usepackage[T1]{fontenc}
\usepackage[version=3]{mhchem}
\usepackage{graphicx}
\usepackage{dcolumn}
\usepackage{bm}
\usepackage{xcolor}
\usepackage{amssymb}
\usepackage{amsmath}
\usepackage{epsfig}
\usepackage{cleveref}
\newcommand{\jk}[1]{\textcolor{black}{#1}}

\begin{document}

\title{Two statistical regimes in the transition to filamentation }

\author{Alexis Gomel}
\affiliation{Group of Applied Physics, University of Geneva, 1205 Geneva, Switzerland}
\affiliation{Institute for Environmental Sciences, University of Geneva, 1205 Geneva, Switzerland}

\author{Geoffrey Gaulier}
\affiliation{Group of Applied Physics, University of Geneva, 1205 Geneva, Switzerland}

\author{Debbie Eeltink}
\affiliation{Group of Applied Physics, University of Geneva, 1205 Geneva, Switzerland}
\affiliation{Institute for Environmental Sciences, University of Geneva, 1205 Geneva, Switzerland}
\affiliation{Present address: Laboratory of Theoretical Physics of Nanosystems, EPFL, Ch-1015 Lausanne,
10 Switzerland}

\author{Maura Brunetti}
\affiliation{Group of Applied Physics, University of Geneva, 1205 Geneva, Switzerland}
\affiliation{Institute for Environmental Sciences, University of Geneva, 1205 Geneva, Switzerland }

\author{Jérôme Kasparian\,\orcidlink{0000-0003-2398-3882}}
\email{jerome.kasparian@unige.ch}
\affiliation{Group of Applied Physics, University of Geneva, 1205 Geneva, Switzerland}
\affiliation{Institute for Environmental Sciences, University of Geneva, 1205 Geneva, Switzerland}

\begin{abstract}
We experimentally investigate fluctuations in the spectrum of ultrashort laser pulses propagating in air, close to the critical power for filamentation. Increasing the laser peak power broadens the spectrum while the beam approaches the filamentation regime. We identify two regimes for this transition: In the center of the spectrum, the output spectral intensity increases continuously. In contrast, on the edges of the spectrum the transition implies a bimodal probability distribution function for intermediate incident pulse energies, where a high-intensity mode appears and grows at the expense of the original low-intensity mode. We argue that this dual behavior prevents the definition of a univoquial threshold for filamentation, shedding a new light on the long-standing lack of explicit definition of the boundary of the filamentation regime.
\end{abstract}

\maketitle

\section{Introduction}

Filamentation is a self-guided propagation regime typical of ultrashort, high-power laser pulses~\cite{BraunCCM1994,Chin2007a,CouaiM2007,BergeSNKW2007}.
Beyond a critical power ($3$~GW in air), the Kerr effect balances diffraction. The beam then self-focuses until higher-order nonlinear defocusing effects like ionization or the saturation of the Kerr effect~\cite{BejotKHLVHFLW2010a} come into play. The resulting dynamical balance gives rise to self-guided light structures known as filaments. Filaments can extend over atmospheric scale distances~\cite{RodriBMKYSSSELHSWW2004,DuranHPMDMFVBDSTCDD2013}, opening the way to various applications from remote sensing~\cite{KaspaRMYSWBFAMSWW2003} to lightning control~\cite{ZhaoDWE1995,ComtoCDGJJKFMMPRVCMPBG2000,Houard2023}, THz generation~\cite{SpranPHK2004,HouarLPTM2008,StepaHPBKW2010,DaiglTHWYCDPC2012, CleriPSCSGLCLOFM2013}, fog clearing~\cite{Schimmel2018}, or the triggering of condensation in sub-saturated atmospheres~\cite{HeninPRSHNVPSKWWW2011,JuLWSWGLCLX2012}.
Filaments show very characteristic and even spectacular features, including long-distance propagation, bright light emission due to both plasma emission on the side and spectral broadening in the forward direction, and noise associated to the shockwave~\cite{YuMKSGFBW2003,Lahav2014,Jhajj2014}. Together, these features give rise to the intuitive notion of a qualitatively distinct propagation regime. However, defining a clear threshold for the filamentation regime turns out to be difficult~\cite{ChinCKKT2008}, especially in focused beams where linear propagation can be sufficient to reach the ionization threshold~\cite{Lim2015}, and even produce a denser plasma than Kerr self-focusing does~\cite{reyes_transition_2018}. Similarly, along the propagation axis, defining the onset and end of filaments, and therefore their length, requires somewhat arbitrary choices.

Here, we statistically investigate the transition to filamentation. Shot-to-shot fluctuations of the spectral amplitudes of a high-power laser beam, whether filamenting or not, are known to display characteristic probability distribution functions (PDF): A regular probability distribution is observed in the center of the spectrum, while on its edge the distribution is long-tailed (optical rogue wave \cite{SolliRKJ2007})~\cite{KaspaBWD2009}. We extend this analysis to the sub-filamenting regime and to the transition to filamentation, in order to characterize how the statistics evolves when the incident power progressively reaches the critical power for filamentation. 
The evolution of spectral intensity PDFs turns out to display typical and contrasted signatures on the edges and in the center of the spectrum, respectively, suggesting that transition to filamentation cannot be characterized as a whole. By doing so, we suggest that the ambiguities in the definition of the \jk{boundaries} of the filamentation regime are intrinsic to the physical process rather than instrumental or conceptual limitations, shedding a new light on the long-standing lack of explicit definition of th\jk{is} boundary.

\section{Experimental setup}

\begin{figure}[htb]
	\centering
	\includegraphics[width=0.9\linewidth]{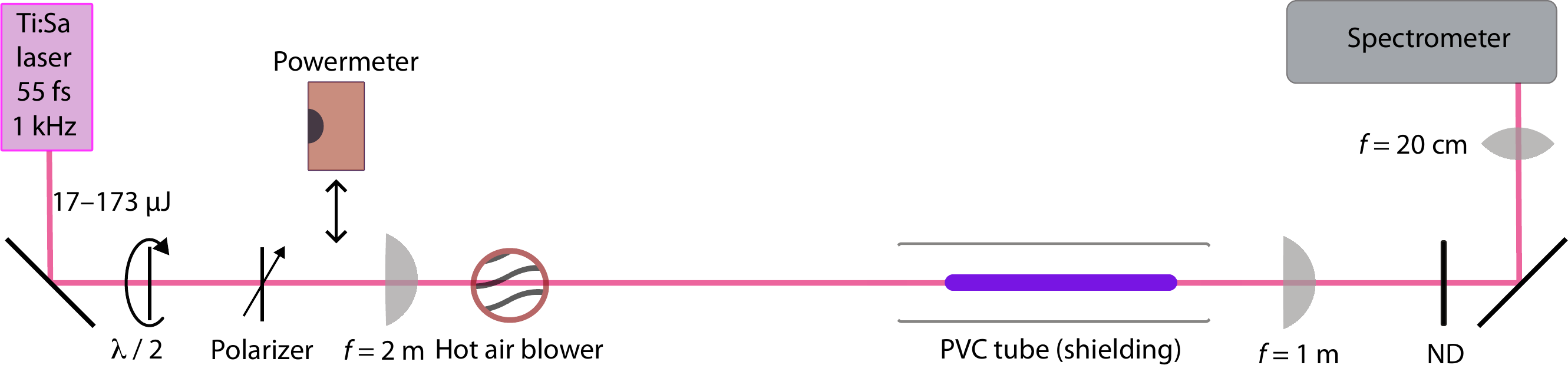}
	\caption{Experimental setup }
	\label{fig: setup_air}
\end{figure}

The experimental setup~(\Cref{fig: setup_air})  relied on a Coherent Astrella laser delivering ultrashort pulses at 1~kHz repetition rate.  
A half-wave plate installed in a motorized rotating stage ensuring $\pm 1^\circ$ precision, followed by a linear polarizer, allowed to adjust the pulse energy with an accuracy of $\sim 10$~$\mu$J. The angle-to-energy calibration was performed before each experimental run with a powermeter (Coherent PM10). \jk{Due to nonlinear chirp that could not be compensated by the compressor, the Fourier-limited 40~fs duration of the pulse was stretched to} $55$~fs (as measured with PulseCheck, APE Berlin), \jk{reaching close to the} filamentation threshold in air within the tuning range of the energy. 
The beam was slightly focused from an initial diameter of 11~mm at $1/e^2$ by an $f = 2$~m BK7 plano-convex lens located approximately at 15~cm  after the polarizer\jk{. We checked that the dispersion of the transmissive optics had a negligible influence on the pulse duration. The beam} subsequently \jk{propagated} through a $13.5$~cm-long turbulent region generated by a \jk{heat gun} (Steinel HL1502 S) transversally \jk{blowing air at $565^{\circ}$C}. $15$--$27.5$~cm downstream of the lens, the beam self-focused and filamentation started if the incident peak power was sufficient. The filamentation region was \jk{shielded against air displacements in the room} by a PVC tube of $18$~cm diameter so as to improve its stability. After the end of \jk{the filamenting} region, the beam was \jk{collimated by an $f = 1$~m lens and focused} by a BK7 lens ($f = 20$~cm) onto the entrance slit of an OceanOptics USB2000 spectrometer providing \jk{$\lesssim$}0.5~nm spectral resolution over the \jk{spectral range of the measurements}. 
A neutral density \jk{(ND)} filter prevented the spectrometer from saturating.
The spectrum was measured and recorded independently for each laser shot. For each incident pulse energy, $5000$ individual spectra were recorded.

\jk{W}e characterized the shot-to-shot fluctuations by recording the PDF of the spectral intensity as a 100-bin histogram, using identical bins in all conditions to facilitate comparison. 
In the characterization of the PDF, we use among others Hogg's unbiased kurtosis estimator~\cite{Kim2004}, defined as 
	\begin{equation}        	   
		Hg=\frac{U_{0.05}-L_{0.05}}{U_{0.5}-L_{0.5}}-2.59
		\label{eq: Hogg}
	\end{equation}
where $U_{m}$ and $L_{m}$ refer to the mean of the upper and lower $m$ quantiles, respectively. This means that $U_{m}=\frac{1}{m}\int_{1-m}^1 F^{-1}(y)dy$ and $L_{m}=\frac{1}{m}\int_{0}^m F^{-1}(y)dy$, where $F$ is the cumulative distribution function. For a Gaussian  PDF, the Hogg estimator is 0.

\section{Results and discussion}

\begin{figure}[htb]
	\centering
	\includegraphics[width=\linewidth]{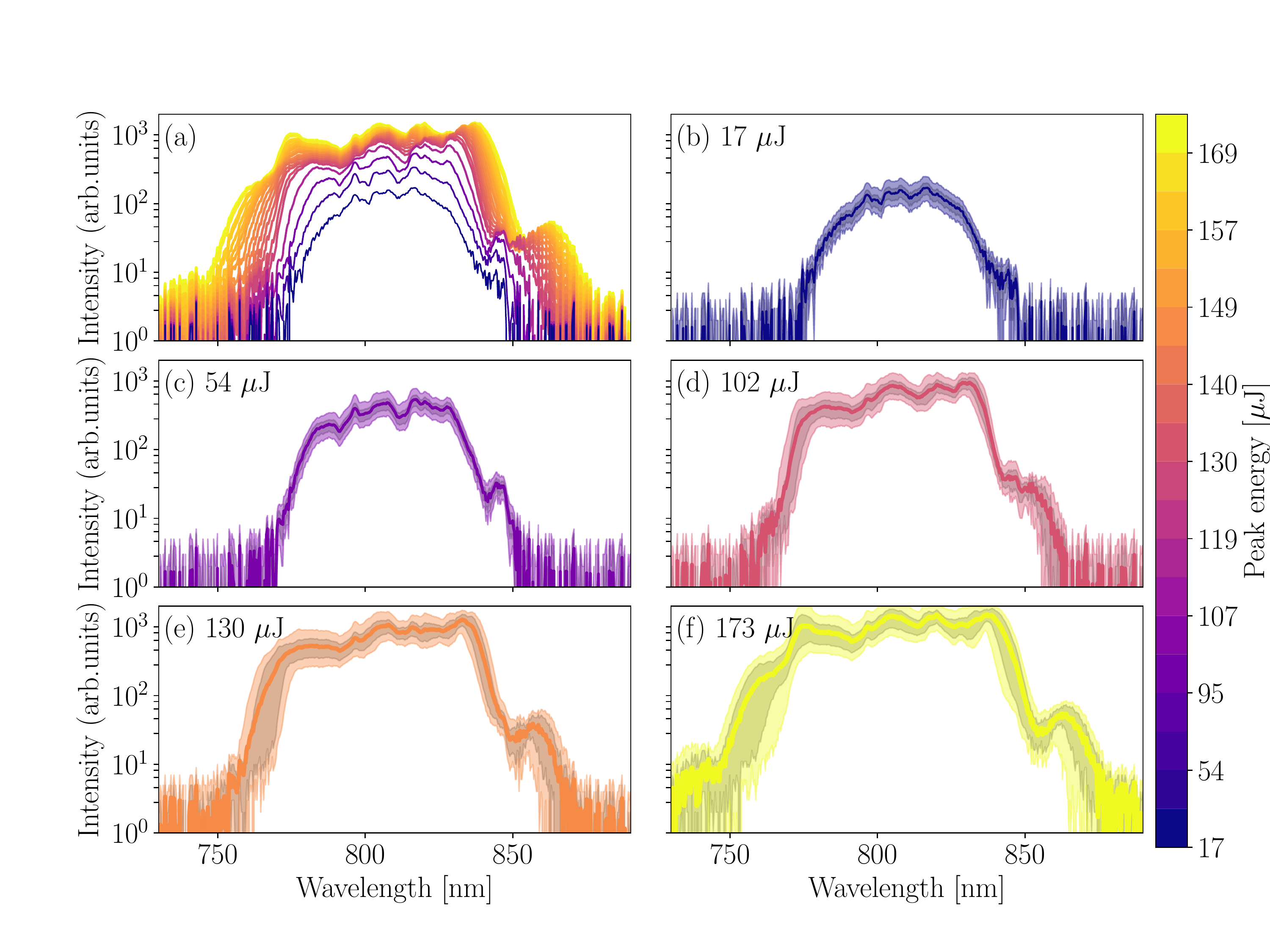}
	\caption{Evolution of the spectrum for increasing peak energy. (a) Average spectra; (b)-(f) Intensity fluctuations for incident pulse energies of 0.017, 0.054, 0.102, 0.130, and 0.173~mJ, respectively. In each panel, the solid thick line is the average intensity, the dark shaded area marks range between the 20$^{th}$ and the 80$^{th}$ percentiles and the light shaded area displays the range between the 5$^{th}$ and the 95$^{th}$ percentiles. }
	\label{fig: spectra examples}
\end{figure}

  Figure~\ref{fig: spectra examples}a displays the \jk{typical} evolution of the \jk{measured} spectr\jk{um}, for incident energies ranging from $0.017$ to $0.173$~mJ/pulse. 
 As is typical for self-phase modulation (SPM)~\cite{Boyd_2007, KaspaBWD2009}, the spectral broadening is characterized by an oscillatory plateau. \jk{Higher incident pulse energies correspond to a broader as well as a wider plateau.}
 
\jk{The turbulence imposed to the beam propagation path, together with } the pre-filamentation fluctuations of the laser itself \jk{in terms of both energy and initial beam profile,} as well as the nonlinear transformation of the spectrum performed by the spectral broadening in the high-intensity region of propagation \jk{result in strong shot-to-shot fluctuations of the output spectrum, as shown by the shaded areas of Fig.~\ref{fig: spectra examples}(b--f).}
The dark (\jk{respectively lighter}) shaded area encompasses \jk{the} 20$^{th}$ \jk{--} 80$^{th}$\jk{(resp. 5$^{th}$ \jk{--} 95$^{th}$) percentile range}.
Shot-to-shot fluctuations of the spectrum \jk{appear} essentially vertical (i.e., in intensity) in the \jk{plateau}, and essentially horizontal (i.e., in spectral range) \jk{in the case of} its edges.

\begin{figure}[htb!]
	\centering
	    \includegraphics[width=0.49\linewidth]{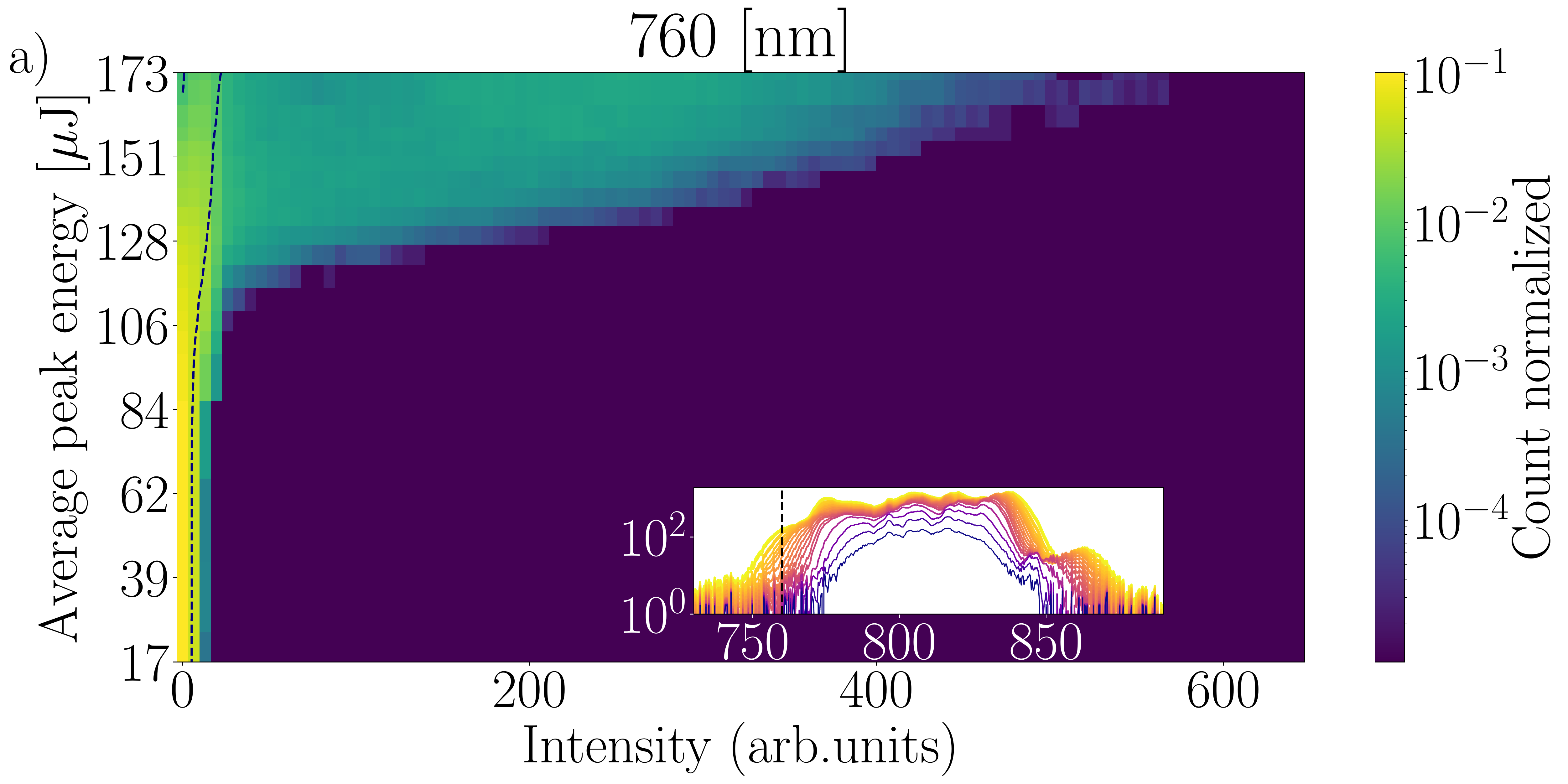}
		\includegraphics[width=0.49\linewidth]{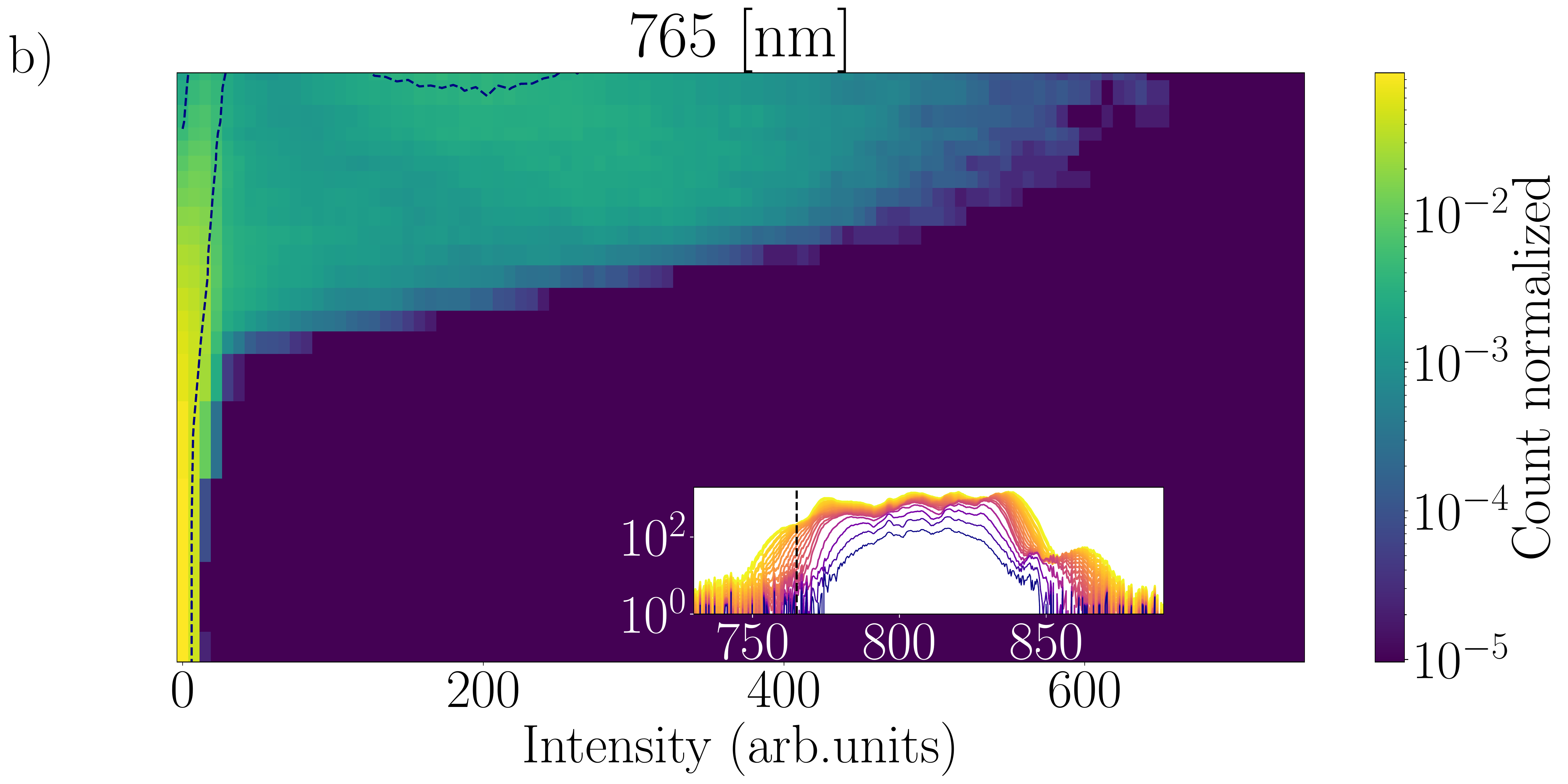}
		\includegraphics[width=0.49\linewidth]{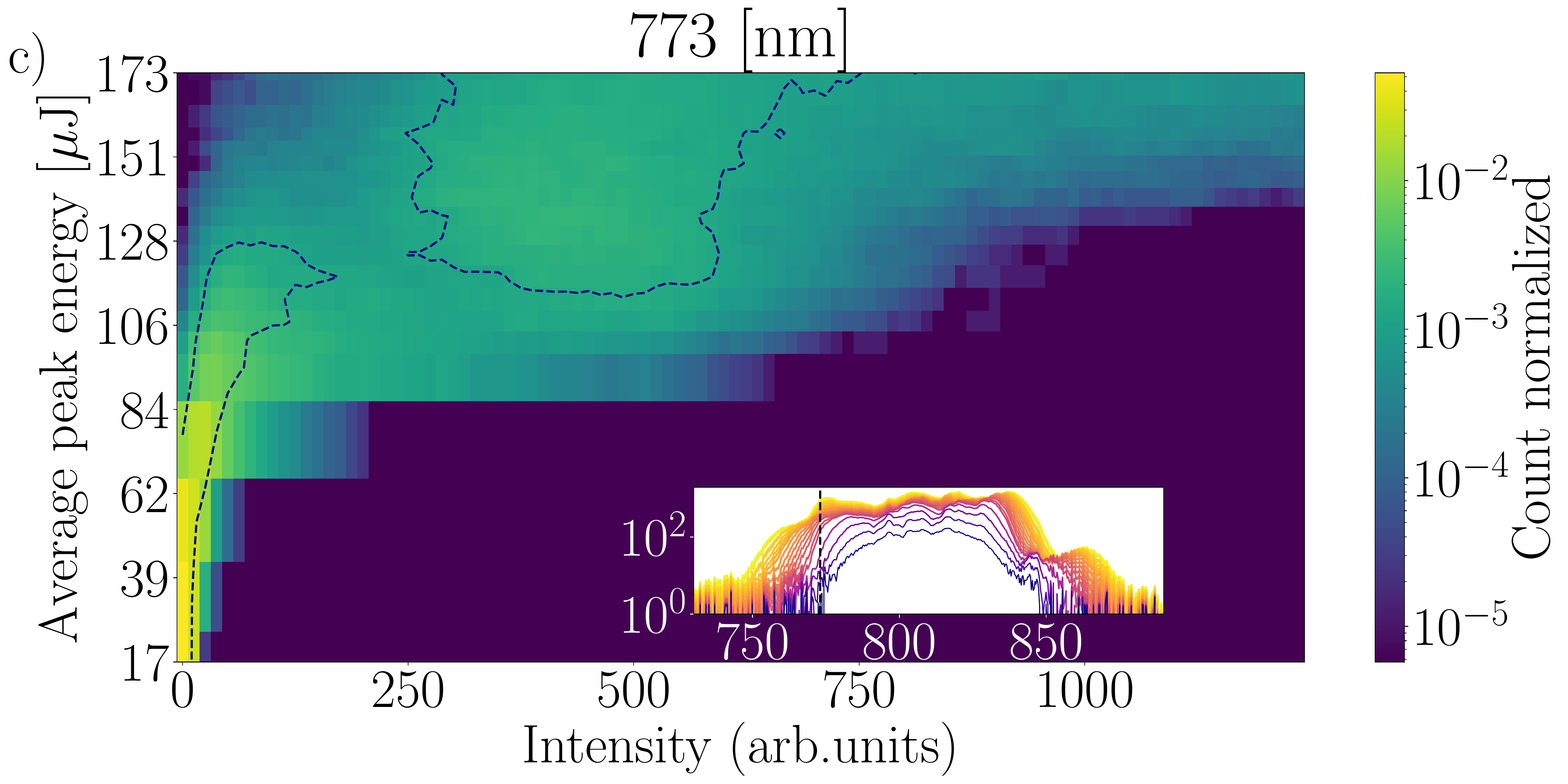}
		\includegraphics[width=0.49\linewidth]{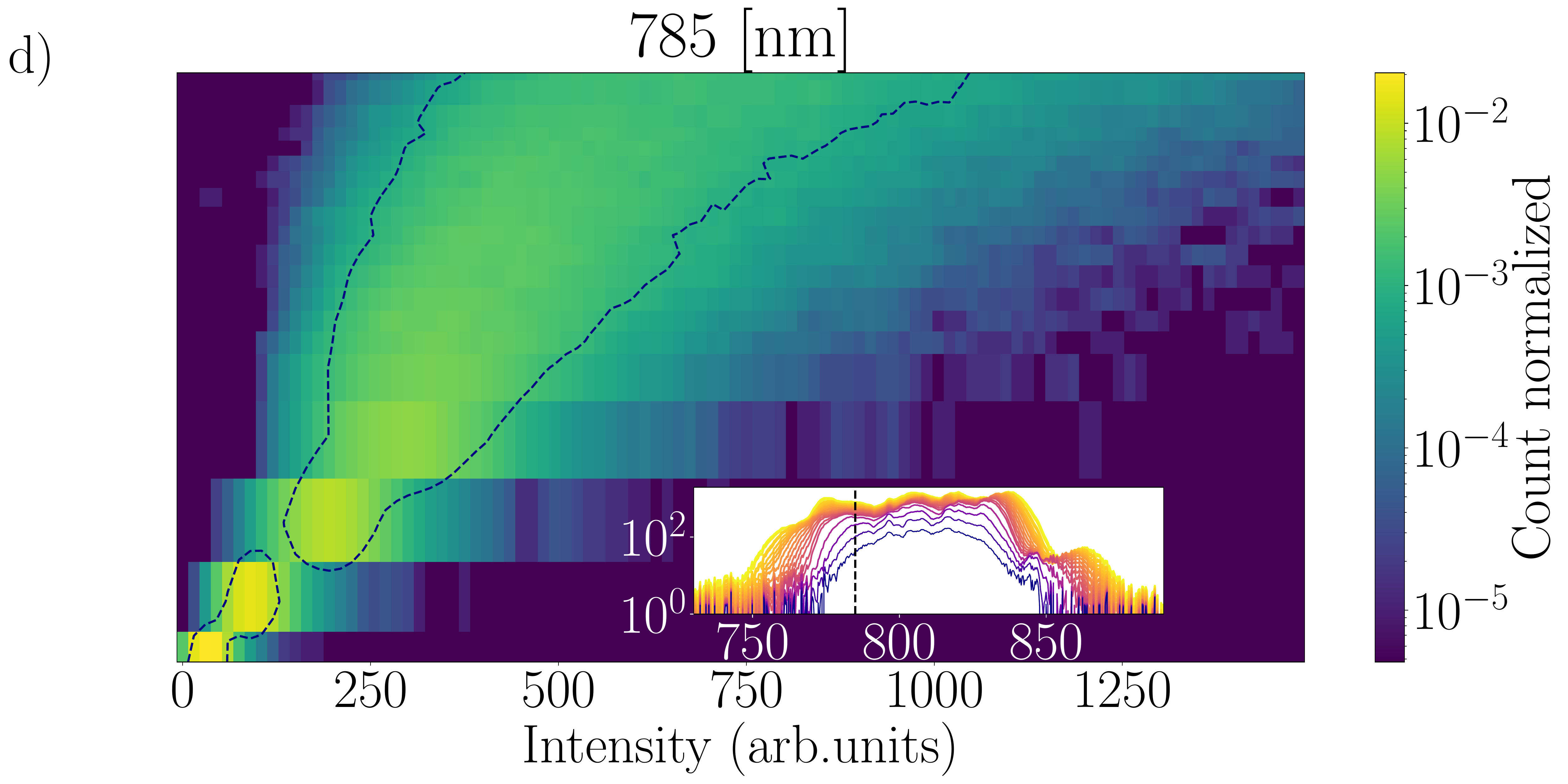}
		\includegraphics[width=0.49\linewidth]{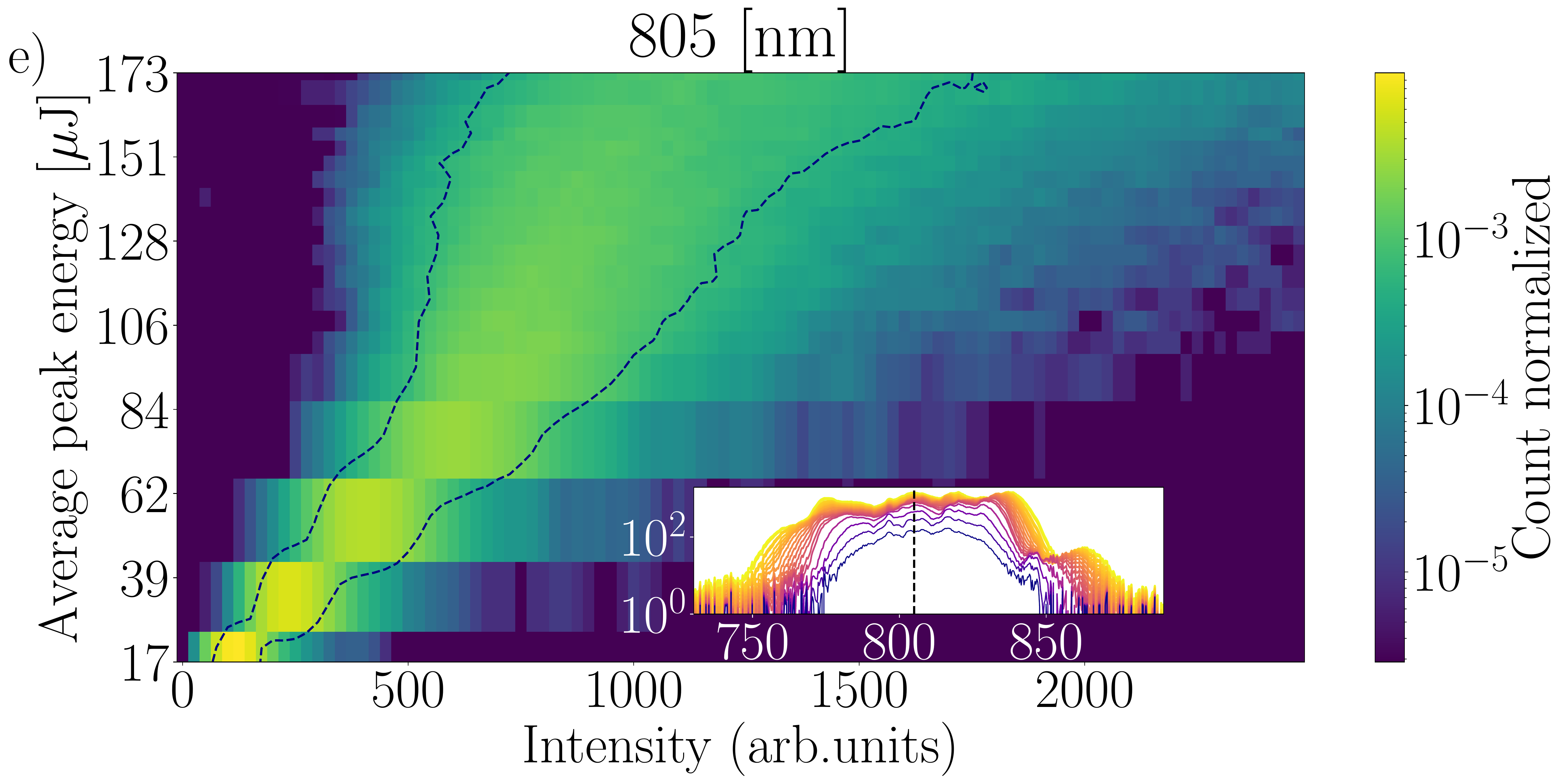}
	    \includegraphics[width=0.49\linewidth]{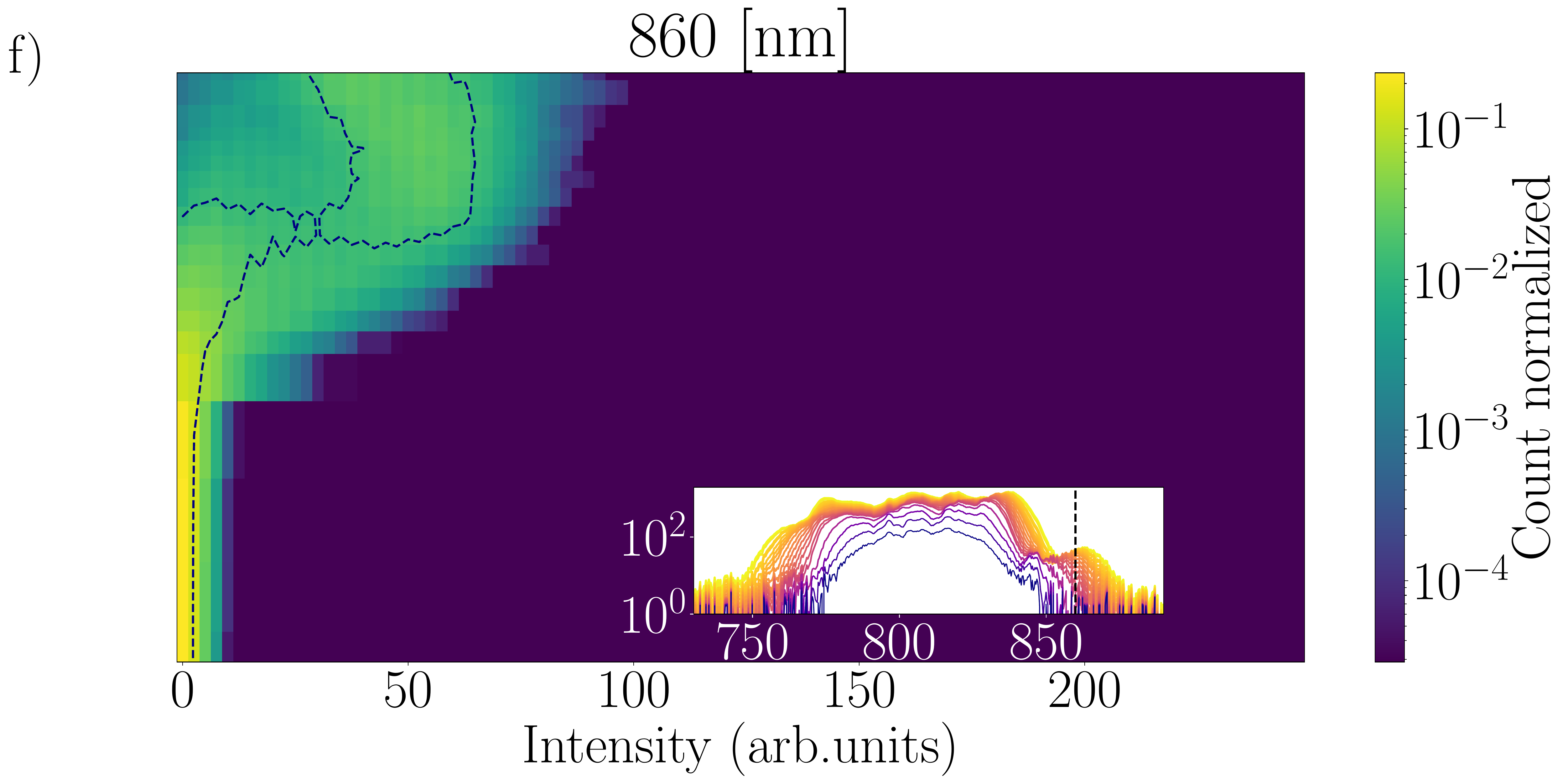}
	\caption{Evolution of the histogram of the spectral intensity at \jk{(a) 760~nm (continuous shift of the distribution mode), (b) 765~nm (bimodal transition), (c) 773~nm (intermediate regime), (d) 785~nm and (e) 805~nm (smooth offset of the peak), and (f) 860~nm (bimodal transition). Each histogram is normalised to 1 and shares the same color scale. Insets display the position of the wavelength in the spectrum (See also \Cref{fig: spectra examples}a) }}
	\label{fig: two_examples}
\end{figure}

\jk{T}he shot-to-shot fluctuations \jk{have an} order of magnitude \jk{comparable to} the mean spectral intensity all over the spectrum. In particular, above the threshold for appreciable spectral broadening, fluctuations reach a factor of 4, ranging from the non-filamenting to the filamenting regimes even at average incident pulse energies up to the critical power (173~$\mu$J, $P$ = 3.15~GW~$\sim P_\textrm{cr}$) where filamentation visually seems to be \jk{already} well established. Such behavior randomly alternating filamenting (spectrally broadened) and non-filamenting pulses is consistent with the previously observed effect of turbulence on filamentation~\cite{AckerMKYSW2006}.

\begin{figure}[htb]
	\centering
	\includegraphics[width=0.8\linewidth]{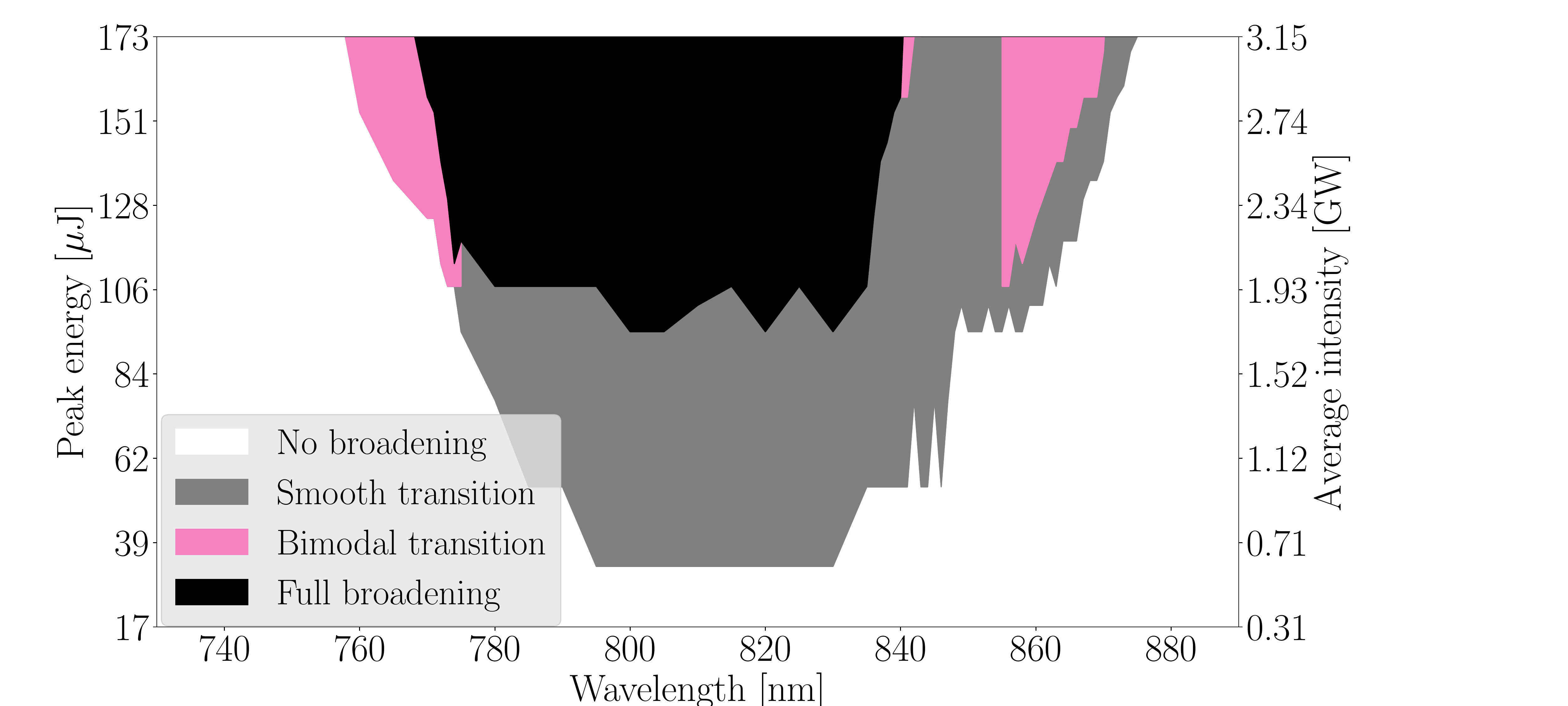}\\
	\includegraphics[width=0.8\linewidth]{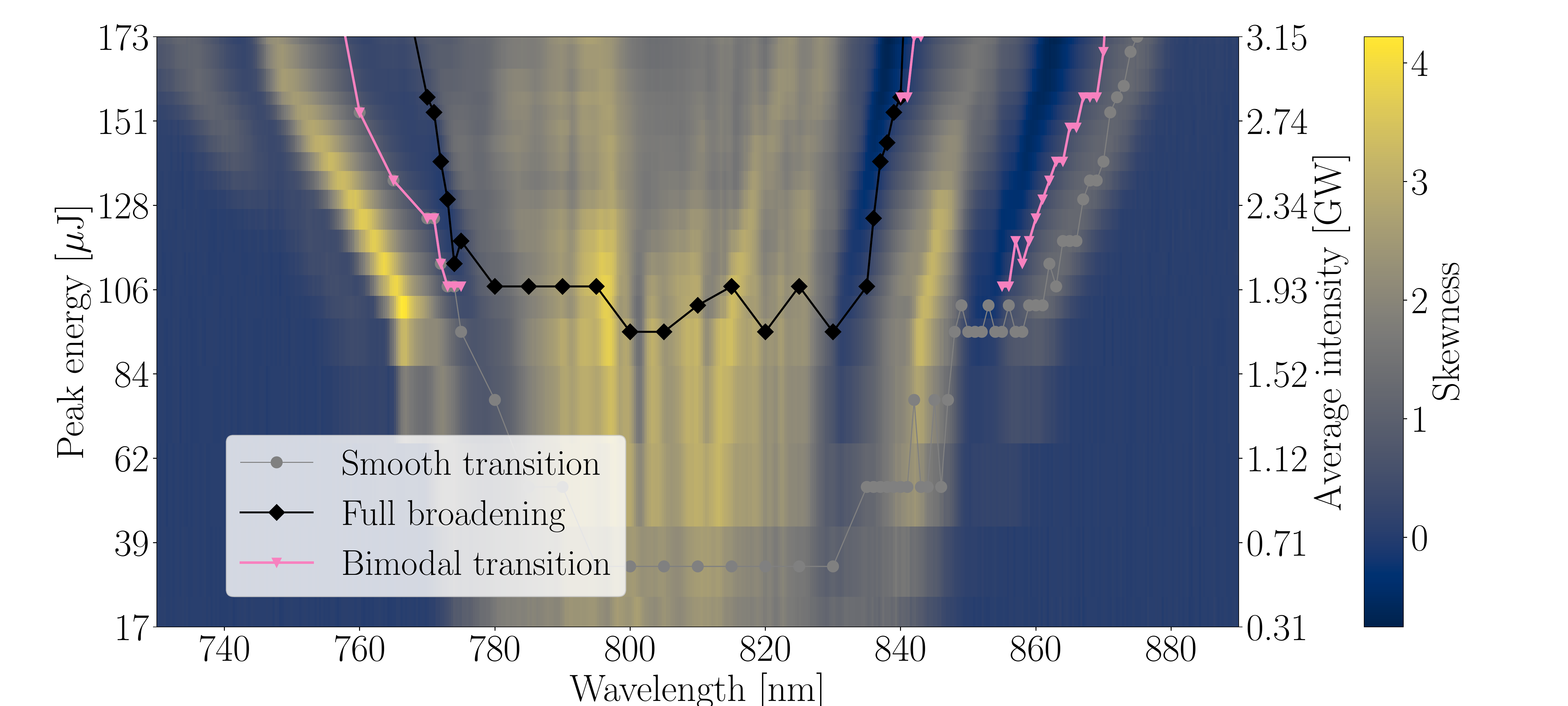}\\
	\includegraphics[width=0.8\linewidth]{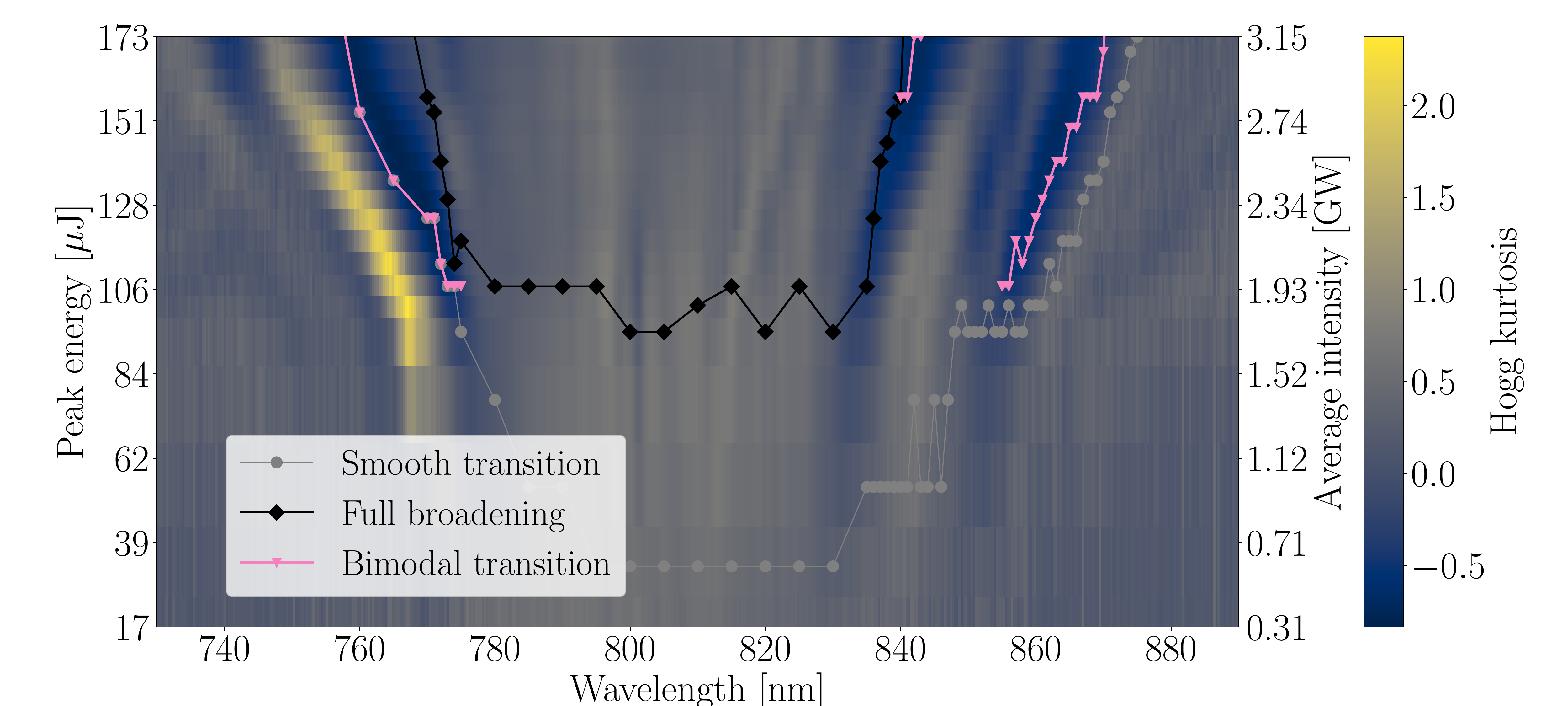}
	\caption{(a) "Phase diagram" displaying the different regimes as a function of wavelength and incident pulse energy, from non-broadened to fully-broadened, with continuous and bimodal transition modes in between. (b, c) Borders of the same diagram overlaid on (b) the skewness and (c) the Hogg's unbiased kurtosis estimator \cite{Kim2004}.}
	\label{fig: bifurcation}
\end{figure}

\Cref{fig: two_examples} \jk{displays the evolution of the PDF of the spectral intensity as a function of the incident energy, for several wavelengths. The behaviors are contrasted. } 
\jk{On the plateau} (805~nm, panel \jk{e}), \jk{the mode of the PDF continuously shifts towards more intense values when} the incident pulse energy \jk{is increased}. This shift becomes faster and faster when moving away from the center of the spectrum \jk{and approaching the side of the plateau} (See 785~nm, panel \jk{d}). \jk{A similar behavior is observed far away from the beam center (760 nm, panel a):} The PDF initially peaks close to zero intensity and \jk{continuously} rises when the \jk{incident pulse energy is sufficient for the} broadening \jk{to reach the considered wavelength}.

\jk{Over a spectral range of $\sim$10~nm on each side of the spectrum, corresponding to both} edges \jk{of the plateau, the transition however occurs in a qualitatively different way} (\jk{765~nm, panel (b) and 773~nm, panel (c), as well as} 860~nm, panel \jk{f}). \jk{T}he low-intensity \jk{mode} progressively vanishes \jk{when the incident pulse energy is increased}, \jk{and a} high-intensity \jk{mode simultaneously} emerges. \jk{For incident energies within the transition, the PDF of the spectral intensity is therefore bimodal}. 

\begin{figure}[htb]
	\centering
	\includegraphics[width=0.8\linewidth]{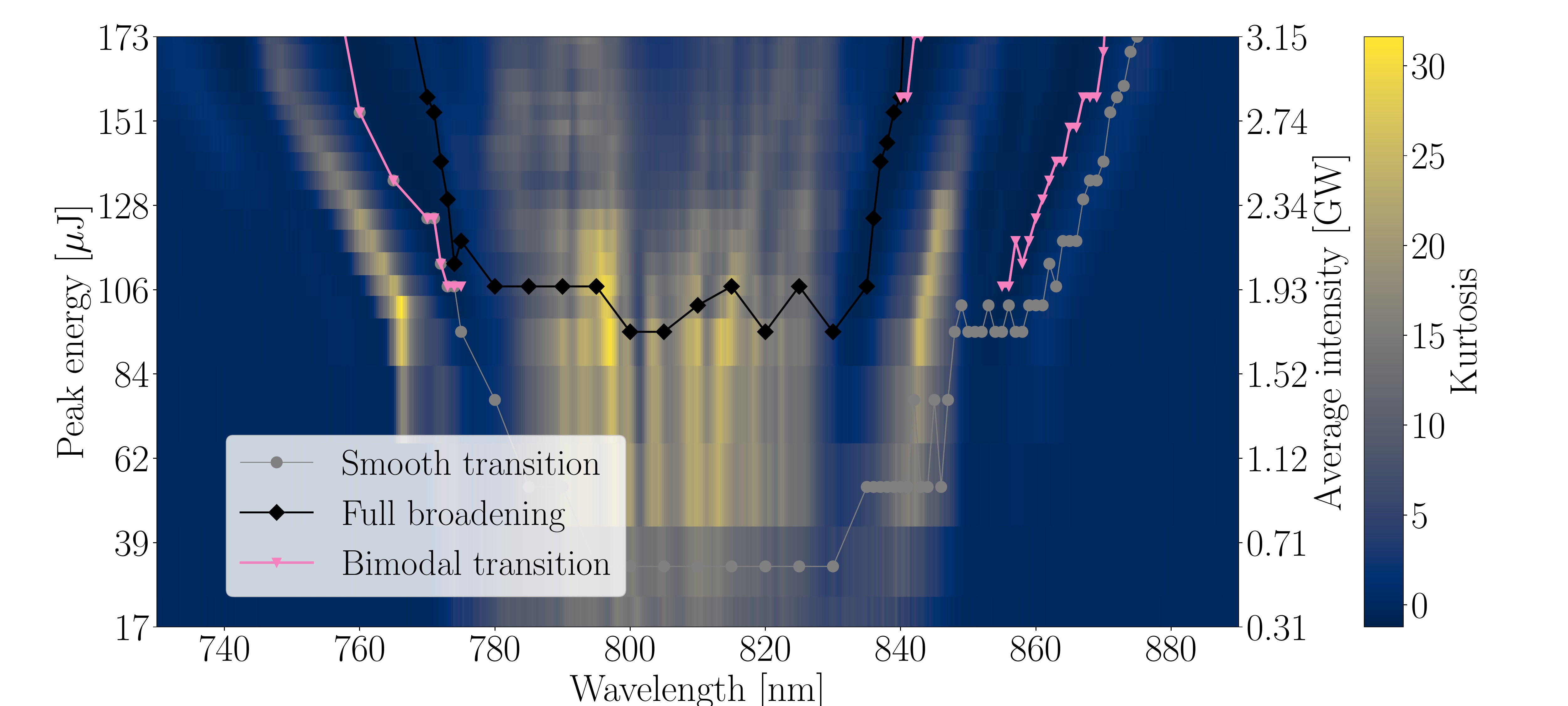}
		\includegraphics[width=0.8\linewidth]{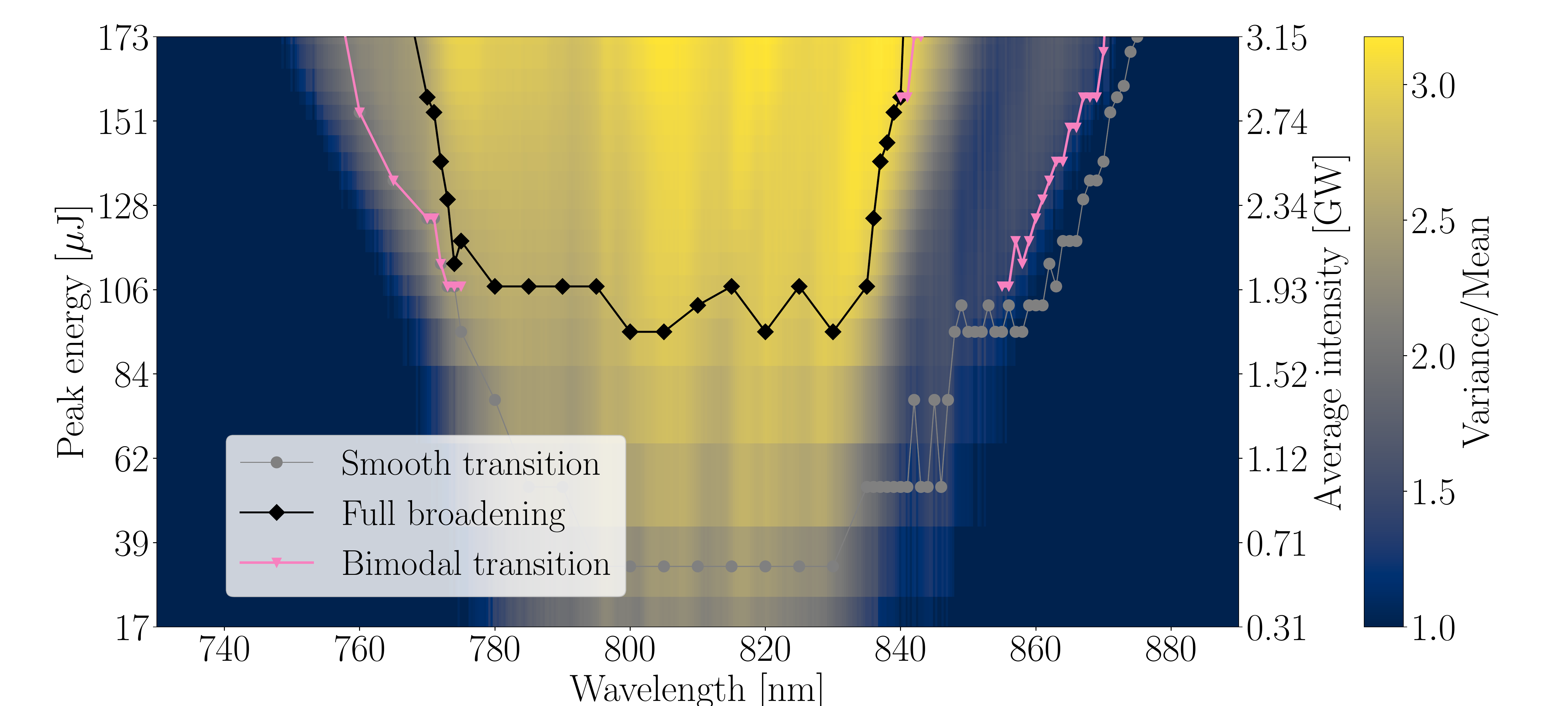}
	\caption{"Phase diagram" displaying the different regimes as a function of wavelength and incident pulse energy, from non-broadened to fully-broadened, with continuous and bimodal transition modes in between overlaid on (a) the kurtosis and (b) the the coefficient of variation (variance/mean).}
	\label{fig:kurtosis}
\end{figure}

\jk{\Cref{fig: bifurcation}a summarises t}he qualitative behaviors \jk{(negligible broadening, fully deployed broadening, continuous transition, and bimodal transition) depending on the} incident pulse energy \jk{and} wavelength. 
\jk{S}kewness (\cref{fig: bifurcation}b) \jk{and} Hogg's unbiased kurtosis estimator $H_g$~\cite{Kim2004} (\cref{fig: bifurcation}c) have low values for bimodal transitions regions \jk{and high values for} continuous transitions. 
This can be understood by considering the corresponding evolutions on \cref{fig: two_examples}. \jk{In the course of} continuous transitions, the PDF tail rises on the high-spectral intensity side, resulting in a heavier tail and \jk{increased} asymmetry. In contrast, \jk{during bimodal transitions,} the secondary mode of the PDF \jk{appears and grows} at the expense of the \jk{PDF} tail\jk{. As a result, the central region of the PDF broadens, so that} the skewness and Hogg estimator \jk{decrease}.
Further statistical moment like the kurtosis and coefficient of variation (i.e.,  the ratio between variance and mean) display qualitatively similar behaviors, as illustrated in \cref{fig:kurtosis}.

\jk{Bimodal transitions occurs on the edge of the plateau, where the shot-to-shot fluctuations of the spectrum mostly consist in horizontal variations} of the cliff delimiting the plateau.
At these wavelengths, the sharp rise of the spectrum on its edge (large $\left|\frac{1}{I}\frac{\textrm{d}I}{\textrm{d}t}\right|$)\jk{, surrounded by two relatively flat regions (the plateau and the low-intensity spectrum side, respectively), implies that t}he spectrum mainly features two ranges of attainable values, \jk{at low spectral intensity on the tail} and \jk{at high intensity} on the plateau, respectively\jk{, resulting in the observed bimodal distribution. T}he intensity \jk{fluctuations} of the plateau itself \jk{are} of second order \jk{here}.  This interpretation also explains why the bimodal behavior is observed on a narrow \jk{spectral} range \jk{close to 840}~nm (See \cref{fig: bifurcation}a), where the plateau \jk{oscillates more}.
In contrast, on the plateau, the fluctuations are mostly vertical. They consist in intensity fluctuations, i.e., \jk{a rise} of frequencies pre-existing in the incident pulse, so that the transition occurs continuously \jk{as} the plateau progressively rises.

These two qualitatively different transition \jk{regimes} coexist \jk{within} the same spectrum\jk{, at frequencies only some nanometers apart and coupled by the nonlinearity of the pulse propagation, and in the course of} the same self-phase modulation process.
The discontinuous\jk{, bimodal} transition to filamentation on the edges of the \jk{plateau} would point to a system where filamenting and non-filamenting regimes are qualitatively distinct and co-exist over some range of input power. In contrast, the smooth, continuous transition \jk{on the plateau in the center} of the spectrum \jk{prevents from defining} a threshold for the filamenting regime. Note that \jk{the latter} spectral range bears most of the pulse energy.

We argue that this dual behavior within the same spectrum during the transition to filamentation intrinsically \jk{implies ambiguities in the} definition of the edges of a filamenting regime that would be qualitatively different from an essentially linear extended focus~\cite{Lim2015}. 

\section{Conclusion}

In summary, we characterized the evolution of intensity fluctuations across the spectrum of an ultrashort beam propagating in air, for a wide range of powers covering below and up to the critical power for filamentation. While on the edges of the spectrum the transition to filamentation is discontinuous and displays a bimodal distribution of spectral intensities around the critical power, it is continuous around the fundamental incident laser wavelength. This dual behavior provides an explanation for the lack of unambiguous definition of the \jk{boundary} of the filamentation regime in experiments or numerical simulations in spite of the intuitive understanding that filamentation qualitatively differs from a more linear propagation regime. 

\smallskip 
\noindent \textbf{Funding.} Swiss National Science Foundation (SNF, grant 200020-175697)

\noindent {\bf Acknowledgements.} Experimental support was provided by Michel Moret. 

\noindent \textbf{Disclosures.} The authors declare no conflicts of interest.

\noindent \jk{\textbf{Data availability.} Data underlying the results presented in this paper are available in Ref.~\cite{Yareta2023}}

\bibliographystyle{apsrev4-1}
\bibliography{biblio_turb}

%

\end{document}